\documentclass[12pt]{article}

\usepackage{amsmath, euscript, amssymb, amsfonts, latexsym}

\newcommand{\Ker}{\mathop{\rm Ker}\nolimits}
\newcommand{\ext}{\mathop{\rm ext}\nolimits}
\newcommand{\Int}{\mathop{\rm Int}\nolimits}
\newcommand{\supp}{\mathop{\rm supp}\nolimits}
\newcommand{\Ima}{\mathop{\rm Ima}\nolimits}

\def\oS{{\Bbb S}}
\def\da{{\rm d}}
\def\ln{\log}
\def\C{{\Bbb C}}
\def\D{{\Bbb D}}
\def\E{{\cal E}}
\def\K{{\cal K}}
\def\R{{\Bbb R}}
\def\U{{\cal U}}
\def\V{{\Bbb V}}

\def\eps{\epsilon}
\def\vep{\varepsilon}
\def\tl{\tilde}
\def\lra{\longrightarrow}
\def\ptl{\partial}
\def\ol{\overline}
\def\wtl{\widetilde}
\def\pr{\prime}

\begin{document}

\baselineskip=20pt

\begin{center}
\Large{Lorentz-covariant ultradistributions, hyperfunctions, and analytic
 functionals}
\end{center}

\smallskip
\begin{center}
{\large M.~A.~Soloviev}
\end{center}

\begin{center}
{\small I. E. Tamm Department of Theoretical Physics, P. N.
Lebedev Physical Institute, Leninsky prosp. 53, Moscow 119991,
Russia, e-mail: soloviev@lpi.ru}
\end{center}

\bigskip

\hfill         {\it Dedicated to Professor V. Ya. Fainberg on the
occasion of 75th birthday} \vspace{1cm}

\bigskip
\begin{center}
\bf{Abstract}
\end{center}

We generalize the theory of Lorentz-covariant distributions to broader
classes of functionals including ultradistributions, hyperfunctions, and
analytic functionals with a tempered growth. We prove that Lorentz-covariant
functionals with essential singularities can be decomposed into
polynomial covariants and establish the possibility of the invariant
decomposition of their carrier cones. We describe the properties of odd
highly singular generalized functions. These results are used to investigate
the vacuum expectation values of nonlocal quantum fields with an arbitrary
high-energy behavior and to extend the spin--statistics theorem to nonlocal
field theory.

\newpage
\vskip2mm

\section{Introduction}

The aim of this work is to extend the theory of Lorentz-covariant
distributions to functionals with singularities of an infinite order. The
theory of Lorentz-covariant distributions plays an important role in the
axiomatic approach~\cite{1},~\cite{2} to quantum field theory (QFT); the main
achievement of this approach is justly considered the derivation of the
spin--statistics relation and the {\sl PCT} symmetry. My interest in highly
singular quantum fields was aroused by Professor V.~Ya.~Fainberg more than 30
years ago, when I was his graduate student. The enthusiasm of those days,
because of the works of Meiman~\cite{3} and Jaffe~\cite{4}, was based on the
hope to solve the nonrenormalizable interaction and zero-charge problems along
this path and to construct a consistent nonlocal field theory. The subsequent
development of gauge theory and superstring theory has shown that some of the
ideas proposed then are still relevant.

Fainberg and Iofa~\cite{5},~\cite{6} first gave a formulation of nonlocal
field theory at the rigor level of the axiomatic approach; the theorem on the
global nature of local commutativity was shown to fail in the case of the
exponential or faster growth of matrix elements of fields in momentum space.
Exactly such a growth, with the exponent proportional to the Planck length,
was later shown for the spectral density in a K\"all\'en--Lehmann-type
representation of string propagators~\cite{7}. Restrictions on the scattering
amplitudes of nonlocally interacting particles~\cite{8}--\cite{10} are
especially interesting because of the AdS/CFT duality (the correspondence
between gravity theories in the anti-de~Sitter space and conformal field
theories on its boundary), which is currently in the focus of attention. In
deriving the scattering matrix from conformal field theory correlators in the
flat limit~\cite{11}, these restrictions can indicate the presence of a
nonlocality.

The use of propagators with nonlocal form factors suppressing ultraviolet
divergences in the Euclidean momentum space has proved efficient in the
Lagrangian formulation of nonlocal QFT~\cite{12} and in the phenomenological
description of strong interactions~\cite{13}. Advanced schemes of this
type involve quantum gravity and are proposed as a phenomenological
alternative to string theory~\cite{14}. In~\cite{15}, it was shown that the
most general distributional framework for constructing local QFT is provided
by the use of the Fourier-symmetric test function space $S^1_1$ and by the
corresponding generalization of microcausality. Subsequently, such a
formulation was developed in great detail in \cite{16},~\cite{17} in terms of
Fourier hyperfunctions. It was found to give a very symmetric relation
between QFT in Minkowski space and Euclidean field theory, which could not be
achieved using the tempered distributions. The theory of the Fourier--Laplace
transformation of functionals on the Gelfand--Shilov spaces
$S^{\beta}_{\alpha}$ constructed in~\cite{18} proved useful in the operator
realization of indefinite-metric gauge models~\cite{19}--\cite{21}, where
singularities have the infrared origin.

A number of theorems on highly singular Lorentz-covariant generalized
functions were established in~\cite{22}. However, the two most interesting,
and also the most difficult, cases were not considered there: the case of
tempered hyperfunctions defined on the space $S^1$, which is universal for
local QFT, and the case of analytic functionals over the space $S^0$, on
which nonlocal fields with an arbitrary high-energy behavior are defined. In
this paper, we completely describe the theory for the functionals of class
$S^{\pr\beta}$, $\beta\geq0$. Although their use in field-theory
constructions has a number of advantages, the test function spaces
$S^{\beta}$ are topologically more complicated than $S^{\beta}_{\alpha}$.
While the latter belong to the well-studied class of DFS spaces (spaces that
are duals of the Fr\'echet--Schwartz spaces), $S^{\beta}$ are not in this
class, which complicates the proof of certain structure theorems for the
functionals defined there.

This complication is overcome using the acyclicity of a sequence of Fr\'echet
spaces whose inductive limit is $S^{\beta}$; we establish the corresponding
theorem in Sec.~2. In the three subsequent sections, we demonstrate the
possibility of splitting the supports of functionals of class $S^{\pr\beta}$
for $\beta>1$ and $\beta=1$ and of their carrier cones for $\beta<1$; proving
this requires diverse arguments. In Sec.~6, these results are used to derive
invariant decompositions of Lorentz-covariant generalized functions; we also
describe the properties of odd invariant functionals with arbitrary
singularities. In Sec.~7, we establish the density of covariant tempered
distributions in the classes of covariant functionals under consideration,
and in Sec.~8, we give the corresponding extension of the representation
through polynomial covariants~\cite{2},~\cite{23}. In Sec.~9, these results
are used to prove the spin--statistics theorem for nonlocal quantum fields;
this proof is an alternative to the one in~\cite{24} using the notion of the
analytic wave-front set.

\section{The spaces $S^{\beta}(O)$ and their topology}

By definition~\cite{25}, the space of test functions $S^{\beta}(\R^n)$ with
the index $\beta\geq0$ consists of infinitely differentiable functions on
$\R^n$ satisfying the inequalities
 \begin{equation}
\bigl|\ptl^{\kappa} f(x)\bigr|\leq
C_NB^{|\kappa|}\kappa^{\beta\kappa}\bigl(1+|x|\bigr)^{-N},
 \label{1}
  \end{equation}
where $\kappa$ ranges the set of multi-indices $\Bbb Z^n$, $N$ ranges the set
$\Bbb N$ of nonnegative positive integers, and the positive constants $C_N$
and $B$ depend on the function $f$. If $\beta>1$, the space $S^{\beta}$
contains functions with a compact support. The space $S^1$ consists of
functions that are analytic in a complex neighborhood of $\R^n$. If
$\beta<1$, the space $S^{\beta}$ consists of functions that can be
analytically continued to the whole of $\C^n$, i.e., are entire functions.
The elements of the dual space $S^{\pr\beta}$ are called ultradistributions
of class $\{\kappa^{\beta\kappa}\}$ (and of tempered growth) in the first
case, hyperfunctions in the second case, and analytic functionals (also of
tempered growth) in the third case. Instead of specifying the topology, the
convergence of sequences in $S^{\beta}$ was defined in~\cite{25}. Namely,
the norms
 \begin{equation}
\|f\|_{B,N}=\sup_{\kappa,x}\bigl(1+|x|\bigr)^N
\frac{\bigl|\ptl^{\kappa} f(x)\bigr|}{B^{|\kappa|}\kappa^{\beta\kappa}}
 \label{2}
  \end{equation}
are associated with inequalities~\thetag{1}, and a sequence $f_{\nu}\in
S^{\beta}$ is said to converge to zero if there exists $B$ such that
$\|f_{\nu}\|_{B,N}\to0$ for any $N$. We show that this definition is entirely
consistent with the natural topologization of $S^{\beta}$ by taking the
projective limit as $N\to\infty$ and the inductive limit as $B\to\infty$.
Another addition to the theory in~\cite{25} that we need in what follows
consists in using similar spaces over open sets in $\R^n$.

{\bf Definition 1.}
Let $O$ be a nonempty open set in $\R^n$. Then $S^{\beta,B,N}(O)$ denotes the
normalized space of infinitely differentiable functions on $O$ with the norm
$\|f\|_{O,B,N}$ defined similarly to Eq.~\thetag{2}, but with the $\sup$
operation taken over $x\in O$. Also, $S^{\beta,B}(O)$ denotes the
intersection $\bigcap_NS^{\beta,B,N}(O)$ endowed with the projective
topology, and $S^{\beta}(O)$ denotes the union $\bigcup_BS^{\beta,B}(O)$
endowed with the inductive topology.

It is easy to verify that the space $S^{\beta,B,N}(O)$ is complete and hence
Banach. Because projective limits inherit the completeness property,
$S^{\beta,B}(O)$ is a Fr\'echet space. To prove the completeness of
$S^{\beta}(O)$, we use a sufficient condition given by Palamodov~\cite{26}.

{\bf Theorem 1.}
{\it The injective sequence of Fr\'echet spaces $S^{\beta,B}(O)$ is acyclic.}

{\bf Proof.}
Let $\U_B$ be a neighborhood of the origin in $S^{\beta,B}(O)$ specified by
$\|f\|_{O,B,0}<1/2$. Obviously, $\U_{B_0}\subset\U_B$ for any $B>B_0$. In
accordance with Theorem~6.1 in~\cite{26}, it suffices to verify that the
topology induced on $\U_{B_0}$ from $S^{\beta,B}(O)$, $B>B_0$, is independent
of $B$. We assume that $f_0\in\U_{B_0}$ and let ${\cal V}_{B,N,\eps}$ denote
the trace on $\U_{B_0}$ of the neighborhood of the function $f_0$ in
$S^{\beta,B}(O)$ given by $\|f-f_0\|_{O,B,N}<\eps$; we show that for
any $B>B_1>B_0$ and any $N_1$ and $\eps_1$, there exist numbers $N$ and
$\eps$ such that ${\cal V}_{B,N,\eps}\subset{\cal V}_{B_1,N_1,\eps_1}$. This
then implies that the topology induced on $\U_{B_0}$ by that of
$S^{\beta,B}(O)$ is not weaker than the topology induced by that of
$S^{\beta,B_1}(O)$; the converse is obvious. In what follows, we set
$\beta=0$ for simplicity (formulas for the general case only differ by
inessential factors). If $f\in{\cal V}_{B,N,\eps}$, we have two estimates for
the function $f_1=f-f_0$,
\begin{equation}
\bigl|\ptl^{\kappa}f_1(x)\bigr|<B_0^{|\kappa|},\qquad
\bigl|\ptl^{\kappa}f_1(x)\bigr|<\eps B^{|\kappa|}\bigl(1+|x|\bigr)^{-N},\quad
x\in O.
 \label{3}
  \end{equation}
We must show that for properly chosen $N$ and $\eps$, this implies that
 \begin{equation}
\bigl|\ptl^{\kappa}f_1(x)\bigr|<
\eps_1B_1^{|\kappa|}\bigl(1+|x|\bigr)^{-N_1},\quad x\in O.
\label{4}
  \end{equation}
We introduce the notation $\vep=\eps\bigl(1+|x|\bigr)^{-N}$ and $\vep_1=
\eps_1\bigl(1+|x|\bigr)^{-N_1}$ and define the number $Q(x)$ by the equation
$B_0^Q=\vep_1B_1^Q$. If $x$ is fixed, the first inequality in~\thetag{3}
implies~\thetag{4} for all $|\kappa|\geq Q$. The second inequality
in~\thetag{3} implies~\thetag{4} for $|\kappa|<Q$ provided that
$\vep B^Q\leq\vep_1B_1^Q$. Setting $\vep B^Q=\vep_1B_1^Q$, we obtain
$\vep=\vep_1^A$, where the number $A=\ln(B/B_0)/\ln(B_1/B_0)$ is independent
of $x$. Therefore, the required implication follows if we take $\eps\leq
\eps_1^A$ and $N\geq AN_1$, which completes the proof.

In accordance with~\cite{26}, the established acyclicity ensures the validity
of the following statements.

{\bf Corollary.}
The space $S^{\beta}(O)$ is Hausdorff and complete. The set ${\cal B}\subset
S^{\beta}(O)$ is bounded if and only if it is entirely contained in some
space $S^{\beta,B}(O)$ and is bounded with respect to each of its norms.

It is certainly obvious that $S^{\beta}(O)$ is a Hausdorff space because its
topology majorizes the uniform convergence topology. We also note that the
inductive limit of Fr\'echet spaces is a bornological space; therefore, the
continuity of a linear mapping of $S^{\beta}(O)$ into an arbitrary locally
convex space is equivalent to its boundedness on all bounded sets, which in
turn is equivalent to the sequential continuity~\cite{27}. The role of the
$S^{\beta}(O)$ spaces in localization problems can be seen from the following
simple remark. An ultradistribution $v\in S^{\pr\beta}$ has support in a
compact set $K$ if and only if for any of its neighborhoods $O$, there exists
a functional $\hat v\in S^{\pr\beta}(O)$ such that $(v,f)=(\hat v,f|_O)$ for
all $f\in S^{\beta}$. For $\beta\leq1$, when there are no functions of
compact support among test functions, this can be taken as the basis for
defining the notion of carrier, replacing the notion of support.

The nuclearity of the spaces $S^{\beta,B+}=\bigcap_{\eps>0}S^{\beta,B+\eps}$
(more precisely, of the Fourier-isomorphic spaces $S_{\beta,B+}$) was proved
in~\cite{28}. This implies the nuclearity of $S^{\beta}$ because inductive
limits of denumerable families of spaces inherit this important property
(see Sec.~3.7.4 in~\cite{27}). In turn, the nuclearity of a space together
with its completeness imply that the space is Montel and, in particular,
reflexive (see Chap.~4, Exer.~19, in~\cite{27}). We use these properties of
$S^{\beta}(\R^n)$ in what follows. The spaces $S^{\beta}(O)$ possess these
properties only under certain restrictions on $O$.

\section{The decomposition of ultradistributions}

Let $v$ be a tempered distribution defined on the Schwartz space $S$, and
let the support of $v$ be contained in the union of closed sets $K_1$ and
$K_2$. It is known that a decomposition of the form $v=v_1+v_2$ with
distributions $v_{1,2}\in S'$ supported by $K_{1,2}$ is possible if these
sets are sufficiently regular and are regularly positioned with respect to
each other. The regularity conditions can be precisely formulated using the
Whitney continuation theorem~\cite{29}. These conditions are trivially
satisfied for sets represented as a union of finitely many closed convex
subsets with a nonempty interior. For what follows, it is useful to recall
the proof of the decomposition theorem in this simplest case (which is
sufficient for most applications). We let $K$ be a set in $\R^n$ of the above
form and let $S(K)$ denote the space of infinitely differentiable functions
on its interior $\Int K$ with the property that their derivatives extended
by continuity to the boundaries of the convex constituents coincide with
each other whenever the boundaries have common points; the functions are
also required to be such that the norms
 \begin{equation}
\max_{|\kappa|\leq N}\sup_{x\in K}\bigl(1+|x|\bigr)^N
\bigl|\ptl^{\kappa}f(x)\bigr|
 \label{5}
  \end{equation}
are finite. This space belongs to the class of FS spaces. By the Whitney
theorem, $f$ can be continued to a smooth function on $\R^n$, which implies
that the space of Schwartz distributions supported by $K$ can be identified
with the dual space $S'(K)$ of $S(K)$. The canonical mapping
 \begin{equation}
S(K_1\cup K_2)\lra S(K_1)\oplus S(K_2)
  \label{6}
  \end{equation}
is injective, is continuous, and has a closed image because the coincidence
of convergent sequences $f_{1\nu}$ and $f_{2\nu}$ on the intersection
$K_1\cap K_2$ implies the coincidence of their limits, which thereby
determine an element of the space $S(K_1\cup K_2)$. This space can therefore
be considered a subspace of the sum $S(K_1)\oplus S(K_2)$. This
identification is valid not only algebraically but also topologically because
a closed subspace of a sum of FS spaces is also a FS space, while in
accordance with the open mapping theorem~\cite{27}, a vector space cannot
have two different comparable topologies such that it is a Fr\'echet space in
each of them. Applying the Hahn--Banach theorem, we conclude that any
functional $v\in S'(K_1\cup K_2)$ has a continuous extension to the sum.
Letting $\hat v$ denote it and writing $v(f)=\hat v(f|_{K_1},0)+
\hat v(0,f|_{K_2})$, we obtain the desired decomposition because composing
$\hat v$ with the canonical embeddings $S(K_i)\to S(K_1)\oplus{\cal S}(K_2)$
gives elements of $S'(K_i)$, $i=1,2$.

Conditions for the decomposability of ultradistributions were studied by
Lambert~\cite{30}, and his theorem covers the functionals of
Gelfand--Shilov's class $S^{\pr\beta}_0$. (We recall that the space
$S^{\beta}_0$ consists of those elements of $S^{\beta}$ that have a compact
support; this space is nontrivial for $\beta>1$.) A similar theorem for the
class $S^{\pr\beta}$ is difficult to prove, but in the particular case that
is solely important for the relevant Lorentz-invariant decompositions
considered in what follows, the corresponding statement is a direct
consequence of Lambert's results.

{\bf Theorem 2.}
{\it Let $K_1$ and $K_2$ be closed convex cones in $\R^n$ such that $K_1\cap
K_2= \{0\}$. Any functional $v\in S^{\pr\beta}$ with support in the cone
$K_1\cup K_2$ can be decomposed into a sum of functionals of the same class
$S^{\pr\beta}$ supported by $K_1$ and $K_2$.}

{\bf Proof.}
In accordance with Theorem 5.1.1 in~\cite{30}, the restriction
$v|_{S^{\beta}_0}$ can be represented as a sum $v_1+v_2$, where the
respective functionals $v_1$ and $v_2$ are supported by $K_1$ and $K_2$. We
need only verify that they have a continuous extension to $S^{\beta}$,
which is then necessarily unique, because $S^{\beta}_0$ is dense in
$S^{\beta}$. Let $\chi$ be an arbitrary function in $S^{\beta}_0$ that is
identically equal to~1 on the ball $U=\bigl\{x:|x|<1\bigr\}$, and let
$\chi_{1,2}$ be multipliers for $S^{\beta}$ (and hence for $S^{\beta}_0$)
that are equal to~1 in a neighborhood of $K_{1,2}\setminus U$ and are equal
to zero in a neighborhood of $K_{2,1}\setminus U$. Because $S^{\beta}_0$ is
an algebra with respect to multiplication, it follows that for all
$f\in S^{\beta}_0$, we have
 \begin{equation}
v_{1,2}(f)=v_{1,2}(\chi f)+v\bigl(\chi_{1,2}(1-\chi)f\bigr).
  \label{7}
  \end{equation}
It remains to note that multiplying by $\chi$ continuously maps $S^{\beta}$
into $S^{\beta}_0$ and the right-hand side of~\thetag{7} is therefore defined
as a continuous functional on $S^{\beta}$.

\section{The decomposition of hyperfunctions}

The regularity conditions for sets established in~\cite{30}, which guarantee
the possibility of the splitting, become progressively weaker as the
functionals become more singular, i.e., as the index $\beta$ of the test
function space decreases. In the class $S^{\pr1}$, the decomposition of a
functional supported by $K_1\cup K_2$ is already possible for any compact
sets $K_1$ and $K_2$. However, the very definition of support is
different in this case because the test functions are analytic. We let $A(K)$
denote the space of functions that are analytic in a complex neighborhood
(depending on a chosen function) of a compact set $K\subset\R^n$ and recall that
it is a DFS space when endowed with the natural topology~\cite{31}.
Directly applying the Taylor formula shows that $S^1$ is continuously
embedded in $A(K)$. In accordance with the standard definition
\cite{29},~\cite{31} of the carrier set of an analytic functional, the
compact set $K$ is a carrier set of $v\in S^{\pr1}$ if $v$ has a continuous
extension to $A(K)$. This can be also expressed as $v\in A'(K)$ because $S^1$
is dense in $A(K)$. The decomposition theorem for $v\in A'(K_1\cup K_2)$ can
be easily proved using the same argument as in the previous section because a
closed subspace of a sum of DFS spaces is a DFS space and a generalization of
the open mapping theorem also applies to spaces of this class. A different
proof, using a harmonic regularization of analytic functionals, is given
in~\cite{29} (Sec.~9.1).

In formalizing the notion of support of elements $v\in S^{\pr1}$, it must
be taken into account that some of them can be naturally regarded as
concentrated at infinity. The point is that the inductive limit $\varinjlim
S^1(O_R)$, where $O_R=\bigl\{x\colon |x|>R\bigr\}$ and $R\to\infty$, is a
Hausdorff space and $S^1$ is injectively and continuously
embedded in it (see Proposition~1.19 in~\cite{22}). Therefore, in accordance
with the Hahn--Banach theorem, there exist nonzero functionals in
$S^{\pr1}$ admitting a continuous extension to this space. The radial
compactification $\D^n=\R^n\sqcup\oS^{n-1}_{\infty}$, where
$\oS^{n-1}_{\infty}$ is an $(n{-}1)$-dimensional sphere at infinity, is
used in the theory of Fourier hyperfunctions~\cite{32}. We also apply it in
the case of functionals of class $S^{\pr1}$. We say that a compact set
$\K\in\D^n$ is a carrier of $v\in S^{\pr1}$ if $v$ can be continuously
extended to the space
 \begin{equation}
S^1(\K)=\varinjlim_{\cal O\supset\K}S^1({\cal O}\cap\R^n),
 \label{8}
 \end{equation}
where ${\cal O}$ ranges open neighborhoods of $\K$ in $\D^n$. If
$\K\subset\R^n$, this definition reduces to the previous one because
inductive limit~\thetag{8} then coincides with $A(\K)$. We recall that
Fourier hyperfunctions compose a space that is the dual of the test function
space $S^1_1$ defined by the inequalities
 \begin{equation}
\bigl|\ptl^{\kappa} f(x)\bigr|\leq
CB^{|\kappa|}\kappa^{\beta\kappa}e^{-|x/A|},
  \label{9}
 \end{equation}
where the constants $A$, $B$, and $C$ depend on $f$. The space $S^1_1(O)$
is a union of Banach spaces $S^{1,B}_{1,A}(O)$ related by compact embeddings
with respect to both indices $A$ and $B$. Therefore, $S^1_1(O)$, as well as
$S^1_1(\K)$, is a DFS space. We next consider the canonical mapping
  \begin{equation}
S^1_1(\K_1\cup\K_2)\lra S^1_1(\K_1)\oplus S^1_1(\K_2)
  \label{10}
 \end{equation}
and use the same simple argument as in the previous section to conclude that
every Fourier hyperfunction with support in $\K_1\cup\K_2$ admits a
decomposition for any compact sets $\K_1,\K_2\subset\D^n$.

The only obstruction to a similar proof of the decomposition theorem for the
functionals of class $S^{\pr1}$ is that the applicability of the open mapping
theorem is no longer obvious. The most general formulation of this
theorem~\cite{33} assumes that the space of values of the mapping is
ultrabornological, i.e., a Hausdorff space representable as the inductive
limit of Fr\'echet spaces. It is by far nonobvious that $S^1(\K)$ with
$\K=\K_1\cup\K_2$ is in this class when it is endowed with the topology
induced by the embedding in $S^1(\K_1)\oplus S^1(\K_2)$. For a deeper
insight into this situation, we must consider the chain of mappings
  \begin{equation}
0\lra S^1(\K_1\cup\K_2)\overset{i}\to{\lra}
S^1(\K_1)\oplus S^1(\K_2)\overset{s}\to{\lra}
S^1(\K_1\cap\K_2)\lra0,
  \label{11}
 \end{equation}
where to a pair of functions $f_{1,2}\in S^1(\K_{1,2})$, $s$ assigns the
difference of their restrictions to $\K_1\cap\K_2$. By Theorem~1.30
in~\cite{22}, the sequence of vector spaces in~\thetag{11} is exact, i.e.,
the kernel of each mapping involved coincides with the image of the preceding
mapping. The only nontrivial point is the exactness in the term
$S^1(\K_1\cap\K_2)$, which means that any element of this space admits a
decomposition into a sum of functions belonging to the spaces $S^1(\K_1)$
and $S^1(\K_2)$ that is established using the H\"ormander $L^2$ estimates in
complete analogy with the corresponding statement for Fourier hyperfunctions.
This implies the existence of supports for the functionals of class
$S^{\pr1}(\R^n)$.

{\bf Theorem 3.}
{\it Each element of the space $S^{\pr1}(\R^n)$ has a unique minimal carrier
in $\D^n$.}

{\bf Proof.}
We first note that if a functional $v\in S^{\pr1}(\R^n)$ has carriers
$\K_1$ and $\K_2$ whose intersection is empty, then $v=0$. This follows
because $S^1$ is dense in $S^1(\K_1\cup\K_2)$ (which is asserted by
Lemma~1.17 in~\cite{22}), and the latter space then contains a function that
is identically equal to zero in a neighborhood of $\K_1$ and is equal to any
chosen element of $S^1$ in a neighborhood of $\K_2$. We next let the carriers
have a nonempty intersection $\K_1\cap \K_2$. To prove the theorem,
it suffices to show that this intersection is a carrier of $v$. Supposing the
converse, we let $\K$ denote the intersection of all carriers
of $v$. If there exists a finite subsystem of the carriers with an empty
intersection, then $v=0$. If there is no such subsystem, then
$\K\neq\varnothing$. Let ${\cal O}$ be a neighborhood of $\K$ in $\D^n$. The
complements of the carriers constitute an open covering of the compact set
$\D^n\setminus{\cal O}$. Choosing a finite subcovering, we conclude
that $\K$ is a minimal carrier.

Going over to the orthogonal complements in the relation $\Ima i=\Ker s$, we
obtain $\Ker i'=\ol{\Ima s'}$, where the primes indicate the conjugate mappings
and the bar means the closure under the weak topology. Because $s$
in~\thetag{11} maps a $\U{\cal F}$ space~\footnote[2]{This class
involves locally convex Hausdorff spaces that can be covered by a denumerable
family of their Fr\'echet subspaces. Any $S^1(\K)$ is in this class because
$S^{1,B}({\cal O}\cap\R^n)$, ${\cal O}\supset \K$, are its Fr\'echet
subspaces and $\K$ has a denumerable fundamental system of neighborhoods in
$\D^n$.} to an ultrabornological space, Grothendieck's formulation of the
open mapping theorem~\cite{34} applies to $s$. In particular, $s$ is a
topological homomorphism. Therefore, the image of $s'$ is weakly closed by
Theorem~4.7.5 in~\cite{27}, and therefore $\Ker i'=\Ima s'$. If
$v\in S^{\pr1}$ has continuous extensions $v_{1,2}$ to $S^1(\K_{1,2})$, then
$i'(v_1,v_2)\bigr|_{S^1}=0$. Because $S^1$ is dense in $S^1(\K_1\cup\K_2)$,
it follows that $i'(v_1,v_2)=0$. Therefore, the functionals $v_{1,2}$ are
restrictions to $S^1(\K_{1,2})$ of elements belonging to $S^{\pr 1}(\K_1\cap\K_2)$,
i.e., there exists a continuous extension of $v$ to $S^1(\K_1\cap\K_2)$,
which completes the proof.

{\bf Theorem 4.}
{\it Let $\K_1$ and $\K_2$ be compact sets in $\D^n$ that are closures of
cones, and let $\K_1\cap\K_2=\{0\}$. Any functional $v\in S^{\pr1}$ with
support in $\K_1\cup\K_2$ admits a decomposition into a sum of functionals of
the same class with supports in the compact sets $\K_1$ and $\K_2$.}

{\bf Proof.}
We let $E$ denote the space $S^1(\K_1)\oplus S^1(\K_2)$ and let
${\cal O}_{1,2}^{\nu}$,  $\nu=1,2,\dots$, be denumerable fundamental systems
of neighborhoods of $\K_{1,2}$ in $\D^n$. The topology of $E$ is identical to
that of the inductive limit of the Fr\'echet spaces
$E_{\nu}=S^{1,\nu}({\cal O}_1^{\nu}\cap\R^n)\oplus
S^{1,\nu}({\cal O}_2^{\nu}\cap\R^n)$. Let $L$ be the image of
$S^1(\K_1\cup\K_2)$ in $E$ endowed with the topology induced by that of $E$.
It suffices to show that this topology coincides with the inductive limit
topology for the family of subspaces $L_{\nu}=E_{\nu}\cap L$. Indeed, the
space $L$ is then ultrabornological (because $L$ being closed in $E$ implies
that $L_{\nu}$ is closed in $E_{\nu}$, i.e., each $L_{\nu}$ is a Fr\'echet
space), which allows applying the Grothendieck theorem~\cite{34}, thereby
completing the proof using the previous argument. The condition for the
coincidence of these two topologies of $L$ follows from the Retakh
theorem~\cite{35} and amounts to the acyclicity of the sequence of quotient
spaces $F_{\nu}=E_{\nu}/L_{\nu}$, which are also Fr\'echet spaces and are
related by canonical injections $F_{\nu}\to F_{\mu}$, $\nu<\mu$. The subspace $L_{\nu}$
is the kernel of the continuous mapping $E_{\nu}\to E/L$; therefore, the
corresponding mapping $E_{\nu}/L_{\nu}\to E/L$ is also continuous. We let $F$
denote the quotient space $E/L$ endowed with the inductive topology relative
to this family of mappings (which is in fact identical to its own topology).
Because sequence~\thetag{11} is exact, it follows that the continuous mapping
$F\to S^1\bigl(\{0\}\bigr)$ is bijective. The space $S^1\bigl(\{0\}\bigr)$
coincides with the ring of germs of analytic functions at $z=0$ and belongs
to the DFS class; therefore, the above bijection is not only an algebraic but
also a topological isomorphism in accordance with the cited
theorem~\cite{34}. Writing $a_{\kappa}=\ptl^{\kappa}f(0)$, we identify the
space $S^1\bigl(\{0\}\bigr)$ with the space of strings $\{a_{\kappa}\}$ of
complex numbers such that $|a_{\kappa}|\leq CB^{|\kappa|}\kappa^{\kappa}$ for
some $C,B>0$. Let $A_{\nu}$ be the Banach space of strings with the norm
$\sup_{\kappa}|a_{\kappa}|/\nu^{|\kappa|}\kappa^{\kappa}$. The proof of
Theorem~1 (where $x=0$ must be set in this simplest case) shows that the
injective sequence of spaces $A_{\nu}$ is acyclic. Therefore, the sequence
$F_{\nu}$ is also acyclic because it is equivalent to the sequence $A_{\nu}$
in view of the above isomorphism. Theorem~4 is thus proved.

\section{The decomposition of functionals in the class $S^{\pr\beta}$,
$\beta<1$}

It was shown in~\cite{22},~\cite{36} that the analytic functionals of class
$S^{\pr\beta}$ (and $S^{\pr\beta}_{\alpha}$), $0<\beta<1$, retain the
angular localizability even though they do not possess supports. Namely, if
arbitrary compact sets in $\D^n$ are replaced with only the closures of cones and
$S^{\beta}(\K)$ is defined similarly to Eq.~\thetag{8}, then the sequence
\begin{equation}
0\lra S^{\beta}(\K_1\cup\K_2)\lra
S^{\beta}(\K_1)\oplus S^{\beta}(\K_2)\lra
S^{\beta}(\K_1\cap\K_2)\lra0
\label{12}
\end{equation}
is exact, which implies that each element of $S^{\pr\beta}$ has the
smallest closed carrier cone (we say that a closed cone in $\R^n$ is a
carrier cone if the compact set obtained by adjoining to this cone the part
of the compact covering $\oS^{n-1}_{\infty}$ cut out by the cone is a
carrier). As
in the case of hyperfunctions, the derivation of this result is based on the
complex-variable representation of the spaces $S^{\beta}(\K)$ associated with
the cones. Namely, let $O$ be the union of an open cone $U\subset\R^n$ and
an $\eps$-neighborhood of the origin. As shown in~\cite{36}, the space
$S^{\beta}(O)$ is isomorphic to the space of entire functions on $\C^n$
satisfying the inequalities
\begin{equation}
\bigl|f(z)\bigr|\leq C_N\bigl(1+|x|\bigr)^{-N}
\exp\Bigl(|By|^{1/(1-\beta)}+d(Bx,U)^{1/(1-\beta)}\Bigr),
\qquad z=x+iy,
\label{13}
\end{equation}
where the constants $C_N$ and $B$ depend on $f$ and $d(\,\cdot\,,U)$ is the
distance from the point to the cone $U$. If $\K$ is the compact set in $\D^n$
corresponding to a closed cone $K$, then each element of $S^{\beta}(\K)$
belongs to some space $S^{\beta}(O)$, where $O$ is of the above form and
contains $K$. The space $S^{\beta}\bigl(\{0\}\bigr)$ associated with the
degenerate cone $\{0\}$ consists of entire functions with an order of growth
at most $1/(1-\beta)$ and of finite type, i.e., the entire functions
satisfying the condition
\begin{equation}
\bigl|f(z)\bigr|\leq C
\exp\Bigl(|Bx|^{1/(1-\beta)}+|By|^{1/(1-\beta)}\Bigr).
\label{14}
\end{equation}

{\bf Theorem 5.}
{\it Let $K_1$ and $K_2$ be closed cones in $\R^n$ such that $K_1\cap
K_2=\{0\}$.  Any functional $v\in S^{\pr\beta}$, $0\leq\beta<1$, with the
carrier cone $K_1\cup K_2$ admits a decomposition into a sum of functionals
of the same class with the carrier cones $K_1$ and $K_2$.}

{\bf Proof.}
The statement of the theorem follows from the exactness of
sequence~\thetag{12} via an argument completely similar to the proof of
Theorem~4 because $S^{\beta}\bigl(\{0\}\bigr)$ is the injective limit of an
acyclic sequence of Banach spaces, as is $S^1\bigl(\{0\}\bigr)$. This is
based on the possibility of decomposing every element of
$S^{\beta}(K_1\cap K_2)$ into functions belonging to $S^{\beta}(K_1)$ and
$S^{\beta}(K_2)$. It is worth noting that for a nonzero $\beta$ and for the
geometry in question, this possibility is almost obvious if we recall that
the space
$S^{\beta}_{1-\beta}$, $\beta>0$, is nontrivial~\cite{25} and contains a
function $\chi_0$ with the properties
\begin{equation}
\bigl|\chi_0(z)\bigr|\leq C_0
\exp\biggl(-\biggl|\frac{x}{A_0}\biggr|^{1/(1-\beta)}+
|B_0y|^{1/(1-\beta)}\biggr)
\label{15}
\end{equation}
and $\int\chi_0(\xi)\,\da\xi=1$. We let $W_1$ and $W_2$ be open cones such
that $K_{1,2}\setminus\{0\}\subset W_{1,2}$ and $\ol{W}_1\cap\ol{W}_2=\{0\}$.
For all $x\in W_1$ and $\xi\in W_2$, we then have the inequality
$|x-\xi|\geq\theta|x|$, where $\theta>0$. We set $\chi(z)=\int_{W_2}
\chi_0(z-\xi)\,\da\xi$. For any $A>A_0$, there is the obvious estimate
\begin{equation}
\bigl|\chi(z)\bigr|\leq C_A\exp\biggl(
-\biggl|\frac{\theta x}{A}\biggr|^{1/(1-\beta)}+
|B_0y|^{1/(1-\beta)}\biggr),\quad x\in W_1.
\label{16}
\end{equation}
The desired decomposition of functions satisfying restrictions~\thetag{14} is
realized by $f=\chi f+(1-\chi)f$ if we take $A_0<\theta/B$, which is always
possible. Indeed, an estimate of type~\thetag{13} (with a sufficiently large
$B_1$ instead of $B$) is then certainly satisfied for the function $\chi f$
when $x\in W_1$, and if the cone $U\supset K\setminus\{0\}$ is such that
$\ol{U}\setminus\{0\}\subset W_1$, this estimate is also satisfied for
$x\notin W_1$ because it then follows that $d(x,U)\geq\theta'|x|$ for some
$\theta'>0$. Therefore, $\chi f\in S^{\beta}(K_1)$. Similarly,
$(1-\chi)f\in S^{\beta}(K_2)$ because $1-\chi=\int_{\complement W_2}
\chi_0(z-\xi)\,\da\xi$ and the cone $\complement W_2$ is separated from $K_2$
by a finite angular distance. For $\beta=0$, this argument does not apply,
and more sophisticated means must be used, as in the theory of
hyperfunctions. It was proved in~\cite{37} that an analogue of
sequence~\thetag{12}
for the spaces $S^0_{\alpha}$ with $\alpha>1$ is exact. In particular,
elements of $S^0\bigl(\{0\}\bigr)=S^0_{\alpha}\bigl(\{0\}\bigr)$ admit a
decomposition even within this narrower class. Therefore, Theorem~5 is also
valid for $\beta=0$. We note that sequence~\thetag{12} is itself exact for
$\beta=0$, which can be verified using Lemma~1.31 in~\cite{24}, but this
proof is considerably more involved.

{\bf Lemma 1.}
{\it Any functional $v\in S^{\pr\beta}$, $\beta\geq0$, whose carrier is the
origin is given by}
\begin{equation}
v=\sum_{\kappa} c_{\kappa}\ptl^{\kappa}\delta(x),\qquad
\lim_{|\kappa|\to\infty}|\kappa|^{\beta}|c_{\kappa}|^{1/|\kappa|}=0.
\label{17}
\end{equation}

{\bf Proof.}
For $\beta>1$, this statement is the simplest case of Theorem~4.1.1
in~\cite{30}. If $\beta\leq1$, the functions $e_{\kappa}=x^{\kappa}/\kappa!$
obviously constitute an unconditional basis in $S^{\beta}\bigl(\{0\}\bigr)$,
and the functionals $e'_{\kappa}=(-1)^{|\kappa|}\ptl^{\kappa}\delta(x)$
constitute the dual basis of functionals. Therefore, $(v,f)=\sum_{\kappa}
(v,e_{\kappa})(e'_{\kappa},f)$, which converts into representation~\thetag{17}
after the redefinition $c_{\kappa}=(-1)^{|\kappa|}(v,e_{\kappa})$. By the
continuity of $v$ in the topology of $S^{\beta}\bigl(\{0\}\bigr)$, we have
the inequality $\bigl|(v,e_{\kappa})\bigr|\leq C_B\|e_{\kappa}\|_B$ for any
$B$, where $\|e_{\kappa}\|_B=\sup_{\ell}\bigl|\ptl^{\ell}e_{\kappa}(0)\bigr|
B^{-|\ell|}\ell^{-\beta\ell}=B^{-|\kappa|}\kappa^{-\beta\kappa}$, which
implies the above restriction on the coefficients $c_{\kappa}$. Lemma~1 is
proved.

\section{The Lorentz-invariant decomposition}

In considering Lorentz-invariant functionals on the Minkowski space $\R^4$,
we use the notation $\ol\V=\{x\in\R^4\colon x^2=x_0y_0-{\bold x}{\bold y}\geq0\}$,
$\ol\V_+=\{x\in\ol\V\colon x_0\geq0\}$, and $\ol\V_-=\{x\in\ol\V\colon x_0\leq0\}$. We
let $L^{\uparrow}_+$ denote the proper Lorentz group.

{\bf Theorem 6.}
{\it Any Lorentz-invariant functional $v\in S^{\pr\beta}$, $\beta\geq0$, with
the carrier cone $\ol\V$ admits a decomposition into Lorentz-invariant
functionals of the same class with the carrier cones $\ol\V_+$ and
$\ol\V_-$.}

{\bf Proof.}
Let $v=v_+-v_-$ be a decomposition of $v$ into functionals with the forward and
backward carrier cones; this decomposition exists in view of Theorems~2, 4,
and~5 (the minus sign here is convenient in what follows). The invariance of
the decomposition with respect to the subgroup of spatial rotations is
ensured by the transition to the averaged functionals $\bar v_{\pm}$,
\begin{equation}
(\bar v_{\pm},f)\stackrel{{\rm def}}{=}
\biggl(v_{\pm},\int_{R\in SO(3)}f(Rx)\,dR\biggr).
\label{18}
\end{equation}
The measure on the rotation group is taken as normalized to~1. This cannot be
applied to pure Lorentzian transformations (boosts) because these
transformations are noncompact. We let ${\sf N}_j=x_j\ptl_0+x_0\ptl_j$ be the
boost representation generators on the space of functionals. It is obvious
that for $\beta>1$, the functional
\begin{equation}
u={\sf N}_1\bar v_+={\sf N}_1\bar v_-
\label{19}
\end{equation}
is supported by the origin. For $\beta=1$, a similar statement is valid in
view of Theorem~3, and for $\beta<1$, the degenerate cone $\{0\}$ is a
carrier of $v$ because sequence~\thetag{12} is exact (and
$S^{\pr\beta}(\,{\ol\V}_{\pm})$ is Lorentz invariant). To obtain the
desired decomposition, it suffices to find a functional of form~\thetag{17}
that is an $SO(3)$-invariant solution of the equation
\begin{equation}
{\sf N}_1v=u.
\label{20}
\end{equation}
Indeed, if such a solution $v_0$ exists, the functionals
$\bar v_{\pm}-v_0\in S^{\pr\beta}(\,{\ol\V}_{\pm})$ are invariant under
the entire group $L^{\uparrow}_+$ in view of the commutation relations
\begin{equation}
[{\sf N}_j,{\sf M}_{ij}]={\sf N}_i,\quad i\neq j,
\label{21}
\end{equation}
where ${\sf M}_{ij}=x_j\ptl_i-x_i\ptl_j$ are the representation generators of
the three-dimensional rotation group. Let ${\sf C}=-\sum_{i<j}{\sf M}_{ij}$
be the Casimir operator of this group. It follows from Eqs.~\thetag{21} that
\begin{equation}
{\sf N}_1u=0,\qquad{\sf C}u=2u.
\label{22}
\end{equation}
The finite-dimensional spaces $E_n$ consisting of functionals of the form
$\sum_{|\kappa|=n}c_{\kappa}\ptl^{\kappa}\delta(x)$ are invariant under
$L^{\uparrow}_+$. Therefore, the problem is reduced to verifying that inside
each $E_n$, the operator ${\sf N}_1$ maps the subspace $F_n=\bigcap_{i<j}
\Ker{\sf M}_{ij}$ onto the subspace determined by Eqs.~\thetag{22}, which is
denoted by $G_n$. The Fourier transformation takes $E_n$ into the space of
homogeneous $n$th-order polynomials, which we decompose into the direct sum
of $SO(3)$-invariant subspaces,
\begin{equation}
E_n=\bigoplus_{l=0}^np_0^{n-l}P_l,
\label{23}
\end{equation}
where $P_l$ consists of homogeneous $l$th-order polynomials in the variables
$p_1$, $p_2$, and $p_3$ (and where $P_0=\C$). We recall~\cite{38} that each
subspace $P_l$ is in turn a direct sum of the minimal invariant subspaces of
the form $({\bold p}^2)^kH_{l-2k}$, $k=0,1,\dots,[l/2]$, where $H_{l-2k}$
consists of harmonic homogeneous polynomials (i.e., those satisfying the
Laplace equation). On $H_{l-2k}$, the Casimir operator is a multiple of the
unit operator, and the corresponding eigenvalue is equal to $2$ only if the
homogeneity degree $l-2k$ is equal to $1$. Further, only those elements of
$H_1$ that are multiples of $p_1$ satisfy the first condition in~\thetag{22}.
Therefore, the polynomials
\begin{equation}
p_0^{n-2k-1}({\bold p}^2)^kp_1,\quad k=0,1,\dots,\left[\frac{n-1}{2}\right],
\label{24}
\end{equation}
constitute a basis in $G_n$. A basis in $F_n$ is formed by the polynomials
\begin{equation}
p_0^{n-2k}({\bold p}^2)^k,\quad k=0,1,\dots,\left[\frac{n}{2}\right].
\label{25}
\end{equation}
Therefore, $\dim F_n\geq\dim G_n$, and the mapping $F_n\to G_n$ under
consideration is indeed surjective. The occurrence of a one-dimensional
kernel of this mapping for even $n$ corresponds to the obvious ambiguity of
the sought decomposition due to the possibility of adding terms of the form
$\sum_lc_l\square^l\delta(x)$. It remains to show that the solution of
Eq.~\thetag{20} can be chosen such that it satisfies the restriction on the
coefficients in Eq.~\thetag{17}. For even $n$, we restrict the mapping
$F_n\to G_n$ to the linear span of the first $(n/2{-}1)$ polynomials
in~\thetag{25}, thereby specifying the choice of the solution $v_0$. Applying
the generator $p_1\ptl_0+p_0\ptl_1$ to these, we see that for any $n$, the
matrix $(a_{kl})$ of the mapping in the above bases is quasi-diagonal with
the only nonzero elements $a_{kk}=n-2k$ and $a_{k,k+1}=2(k+1)$, $0\leq
k\leq\bigl[(n-1)/2\bigr]$. The inverse matrix is also upper-triangular, the
absolute value of its elements increases monotonically as both indices
simultaneously increase and reaches its maximum at $l=\bigl[(n-1)/2\bigr]$,
where
\begin{equation}
\bigl|a^{(-1)}_{kl}\bigr|=\frac{(n-1)!!}{(n-2k)!!\,(2k)!!}
\label{26}
\end{equation}
for odd $n$. For even $n$, the numerator in~\thetag{26} is replaced with
$(n-2)!!$. The product of the left- and the right-hand sides of the
inequality
\begin{equation}
\frac{(n-1)!!}{(n-2k)!!\,(2k)!!}\leq\frac{n!!}{(n-2k-1)!!\,(2k-1)!!}
\nonumber
\end{equation}
is the binomial coefficient $\binom{n}{2k}$, which is majorized by $2^n$.
Therefore, $\bigl|a_{kl}^{(-1)}\bigr|\leq2^{n/2}$. Next, the decomposition
coefficients of the elements of $G_n$ with respect to basis~\thetag{24} are a subset of the
decomposition coefficients with respect to the monomials $p^{\kappa}$, and
the decomposition coefficients of the elements of $F_n$ with respect to the monomials differ
from the decomposition coefficients with respect to basis~\thetag{25} by
factors that are not greater than $3^{n/2}$. Thus, if $u=\sum_{\kappa}
c_{\kappa}\ptl^{\kappa}\delta(x)$, then the coefficients $c^0_{\kappa}$ of
the above solution $v_0$ to~\thetag{20} satisfy the estimate
$\max_{|\kappa|=n}|c^0_{\kappa}|\leq6^{n/2}\max_{|\kappa|=n}|c_{\kappa}|$,
which completes the proof of Theorem~6.

{\bf Remark.}
The proof presented can be directly extended to the space--time of an
arbitrary dimension $d\geq3$. However, for $d=2$, where the rotation subgroup
is absent, the proof loses its applicability, and a similar theorem is not
true.

Indeed, we consider the distribution $\ptl_+\bigl[\theta(x_+)\ln x_+\bigr]
\delta(x_-)\in S'(\R^2)$, where $x_{\pm}=(x_0\pm x_1)/\sqrt2$ are the
light-cone coordinates. This distribution is not Lorentz-invariant in $\R^2$,
is supported by a ray on the boundary surface of the cone $\ol\V_+^{(2)}$,
and satisfies the equation ${\sf N}_1v=\delta(x)$ (in these variables,
${\sf N}_1=x_+\ptl_+-x_-\ptl_-$). The same equation is satisfied by the
distribution obtained from this one via the reflection $x\to-x$; however,
there is no solution to this equation among functionals~\thetag{17}.
Therefore, the sum
\begin{equation}
\ptl_+\bigl[\theta(x_+)\ln x_+\bigr]\delta(x_-)+
\ptl_+\bigl[\theta(-x_+)\ln|x_+|\bigr]\delta(x_-)
\label{27}
\end{equation}
is an odd Lorentz-invariant distribution, which does not admit an
invariant decomposition in either the Schwartz class $S'$ or the classes
$S^{\pr\beta}$, $\beta\geq0$. This example also shows that not every
Lorentz-invariant distribution in $\R^2\setminus\{0\}$ admits an invariant
extension to the whole of the $\R^2$ space, whereas for $\R^4$, such an extension
always exists (Proposition~3.5 in~\cite{2}). Because the distribution
$\theta(x_+)x_+^{-1}\delta(x_-)$ on $\R^2\setminus\{0\}$ cannot be continued
to a Lorentz-invariant positive measure on $\R^2$, it is necessary to use an
infrared indefinite metric in quantizing the massless scalar field in two-dimensional
space--time~\cite{39}.

{\bf Theorem 7.}
{\it Let $v$ be a Lorentz-invariant functional of class $S^{\pr\beta}$,
$\beta\geq0$, with the carrier cone $\ol\V$. If its Fourier transform vanishes
in a neighborhood of a spacelike point, then $v$ is odd.}

{\bf Proof.}
Let $v=v_+-v_-$ be the invariant decomposition of $v$ into functionals with
the carrier cones $\ol\V_+$ and $\ol\V_-$. Theorem~4 in~\cite{40} shows
that there is a well-defined Laplace transform ${\bold u}_{\pm}(\zeta)=
\bigl(v_{\pm},e^{i(\cdot,\zeta)}\bigr)$ of $v_{\pm}$. The functions
${\bold u}_{\pm}$ are analytic in the domain $\Bbb T_{\pm}=\{\zeta=p+i\eta\colon
\eta\in\V_{\pm}\}$, and their boundary values as $\eta\to0$,
$\eta\in V'_{\pm}$ (where $\ol V^{\,\pr}_{\pm}\setminus\{0\}\subset\V_{\pm}$),
are just the Fourier transforms $u_{\pm}={\cal F}v_{\pm}$. In~\cite{40}, the
classes $S^{\pr0}_{\alpha}\supset S^{\pr\beta}$ were considered, and the
convergence to the boundary values was proved in the topology of
$S^{\pr\alpha}_0={\cal F}(S^{\pr0}_{\alpha})$, but in the present case,
certainly, this convergence also occurs in the topology $S^{\pr}_{\beta}=
{\cal F}(S^{\pr\beta})$, as can be verified using the same argument. In
accordance with the Bargmann--Hall--Wightman theorem \cite{1},~\cite{2}, each
Lorentz-invariant function ${\bold u}_{\pm}$ can be analytically continued to
the extended domain $\Bbb T^{\ext}$ that contains all spacelike points.
This continuation is symmetric with respect to the complex Lorentz
group $L_+(\C)$ and, in particular, with respect to the full reflection
$\zeta\to-\zeta$. Taking the uniqueness property of analytic functions into
account (Sec.~6 in~\cite{41}), we conclude that the above assumption about
the support of $u={\cal F}v$ implies the equality ${\bold u}_+(\zeta)=
{\bold u}_-(\zeta)$, $\zeta\in\Bbb T^{\ext }$. Therefore, for any test
function $g\in S_{\beta}={\cal F}(S^{\beta})$, we have
\begin{equation}
\split (u,g)&=\lim_{\eta\to0,\eta\in V'_+}
\int\bigl(u_+(p+i\eta)-u_-(p-i\eta)\bigr)g(p)\,\da p=
\\
&= \lim_{\eta\to0,\eta\in V'_+}\int\bigl(u_-(-p-i\eta)-
u_+(-p+i\eta)\bigr)g(p)\,\da p=-\bigl(u,g(-\,\cdot\,)\bigr),
\endsplit
\nonumber
\end{equation}
as was to be proved.

We note that Theorem 7 can be strengthened somewhat: in the condition of the
theorem, it suffices to suppose that the functional $v$ has a closed
carrier cone $K\supset\V$ that is different from the entire space. Then
$\ol\V$ is also its carrier cone in view of the existence of the smallest
closed carrier cone and in view of the Lorentz invariance of $v$, because any
spacelike direction can be rotated into the interior of the complement of $K$
by an appropriate Lorentz transformation.

\section{The denseness theorem}

Let $\E$ be a finite-dimensional complex vector space carrying a
representation $T$ of the group $L^{\uparrow}_+$. We let $L(S^{\beta},\E)$
denote the space of continuous linear mappings of $S^{\beta}$ into $\E$ endowed
with the topology of uniform convergence on bounded sets. With a basis fixed in
$\E$, a mapping $w\in L(S^{\beta},\E)$ can be identified with the set of
continuous linear functionals $w^j\in S^{\pr\beta}$ (or, equivalently, with
an element of the space $(\E'\otimes S^{\beta})'$); the number of these
functionals is equal to the dimension of the representation. Because the
space $S^{\beta}$ is Montel, the convergence of a sequence $w_{\nu}$ in the
above topology is equivalent to the condition that each sequence $w^j_{\nu}$
is weakly convergent. A mapping $w\in L(S^{\beta},\E)$ is called a
(vector-valued) Lorentz-covariant generalized function if it satisfies the condition
\begin{equation}
w(f)=T(\Lambda)w(f_\Lambda),\quad
\Lambda\in L^{\uparrow}_+,\quad f\in S^{\beta},
\label{28}
\end{equation}
where the Lorentz group action in the space of test functions is defined in
the standard way as $f_\Lambda(x)=f(\Lambda^{-1}x)$. If $T$ has an odd
valence and realizes a representation of the $SL(2,\C)$ group, which is the
universal covering of $L_+^{\uparrow}$, then a condition similar
to~\thetag{28} implies the identical vanishing $w\equiv0$, which therefore
means that we are in fact dealing with single-valued Lorentz group
representations. In the component notation, condition~\thetag{28} becomes
\begin{equation}
(w^j,f_{\Lambda^{-1}})=\sum_kT^j_k(\Lambda)(w^k,f),\quad
\Lambda\in L^{\uparrow}_+,\quad f\in S^{\beta}.
\label{29}
\end{equation}
Lorentz-covariant generalized functions constitute a closed subspace in
$L(S^{\beta},\E)$, which we endow with the induced topology.

{\bf Theorem 8.}
{\it For any $\beta\geq0$, the space of tempered Lorentz-covariant generalized
functions is dense in the space of covariant distributions of the class
$S^{\pr\beta}$ transforming according to the same Lorentz group
representation.}

{\bf Proof.}
We regularize the ultraviolet behavior of a covariant generalized function
$w$ by multiplying its Fourier transform $u={\cal F}w$ by $\chi(p/M)=
\chi_0(p^2/M^2)$, where $p^2$ is the Lorentz square of the vector $p$ and
$\chi_0(t)$ is an infinitely differentiable function with support in the
interval $(-1,1)$ and identically equal to~1 for $|t|\leq1/2$. The space
$S_{\beta}$ consists of smooth functions $g$ such that
\begin{equation}
\|g\|_{B,N}=\max_{\kappa\leq N}\sup_p\bigl|\ptl^{\kappa}g(p)\bigr|
\exp\left(\left|\frac{p}{B}\right|^{1/\beta}\right)<\infty
\label{30}
\end{equation}
for some $B$ (depending on $g$) and any $N=0,1,\dots$\,. Therefore, it is
obvious that $\chi$ is a multiplier for $S_{\beta}$, and the estimate
\begin{equation}
\left|\ptl^{\kappa}\chi\left(\frac{p}{M}\right)\right|\leq
C_{\kappa}\left|\frac{p}{M^2}\right|^{|\kappa|}
\label{31}
\end{equation}
allows an easy verification that $g\chi(p/M)\to g$ in $S_{\beta}$ as
$M\to\infty$. Therefore, $u^j_M= u^j\chi(p/M)\to u^j$ in $S'_{\beta}$. It
remains to show that the functionals $u^j_M$ admit a continuous extension to
the Schwartz space $S$. Because $S'_{\beta}\subset S'_0={\cal D}'$, we can
use the criterion noted in~\cite{42} (Sec.~2.10.7), in accordance with which
a distribution in ${\cal D}'$ can be continued to $S$ if and only if its
convolution with any test function $g\in{\cal D}$ supported by the unit ball
$|p|<1$ has no worse than a power growth at infinity. We first consider the case of
Lorentz-invariant functionals. The value of the convolution $(u_M*g)$ at a
point $q$ is the value taken by the distribution $u_M$ on the shifted
function $g(p-q)$. We need only consider the shifts along the light-cone
surface because for the other directions, $(u_M*g)(q)$ vanishes for
sufficiently large $|q|$. It can be additionally assumed that $q_2=q_3=0$,
because any vector $q'\in\R^4$ is taken into a certain point $q$ in this
plane by an appropriate spatial rotation $R$ and $(u_M*g)(q')=(u_M*g_R)(q)$,
where $g_R(\,\cdot\,)=g\bigl(R^{-1}(\,\cdot\,)\bigr)$. Finally, we can set
$M=1$ without loss of generality because $S'$ and $S'_{\beta}$ are invariant
under dilations. We now use the light-cone variables $q_{\pm}=(q_0\pm q_1)/\sqrt2$
and set $q_-=0$ and $q_+\to+\infty$ for definiteness. We let $\Lambda$ denote
the Lorentz transformation $p_+\to p_+/q_+$, $p_-\to q_+p_-$ in the plane
$(p_0,p_1)$, which takes  $q$ into a vector with the unit Euclidean norm. In
view of the Lorentz invariance of $u$ and $\chi$, we have
\begin{equation}
(u\chi*g)(q)=(u,g_q),\quad\text{where }g_q(p)=\chi(p)g(q-\Lambda^{-1}p).
\label{32}
\end{equation}
The points of $\supp g_q$ satisfy the inequalities $|p^2|<1$ and
$p_2^2+p_3^2<1$ by construction, and hence $|p_+p_-|<1$. In addition,
$|q_+-q_+p_+|<1$, and consequently $|p_-|<1/(1-1/q_+)$. Therefore, if $q_+$
is sufficiently large, it follows that $\supp g_q$ is contained in a ball of
radius~2, and we have the estimate
\begin{equation}
\bigl|(u,g_q)\bigr|\leq\|u\|_{2,N}\|g_q\|_{2,N},
\label{33}
\end{equation}
where $\|g_q\|_{2,N}=\max_{|\kappa|\leq N}\sup_{|p|\leq2}
\bigl|\ptl^{\kappa}g_q(p)\bigr|$ in accordance with Eq.~\thetag{30} and $N$
has the meaning of the singularity order of the distribution $u$ in the ball
$|p|<2$. The transformation $\Lambda^{-1}$ results in contracting the graph
of $g$ by $q_+$ times with respect to the variable $p_+$; therefore,
\begin{equation}
\sup_p\bigl|\ptl^{\kappa}g(q-\Lambda^{-1}p)\bigr|=
\sup_p\bigl|\ptl^{\kappa}g(\Lambda^{-1}p)\bigr|\leq
C_{\kappa}\|g\|_{1,N}q_+^{|\kappa|},\quad|\kappa|\leq N.
\label{34}
\end{equation}
Together with estimate~\thetag{31}, this gives
\begin{equation}
\|g_q\|_{2,N}\leq C_N\|g\|_{1,N}\bigl(1+|q|\bigr)^N.
\label{35}
\end{equation}
We therefore conclude that the behavior of $(u_M*g)(q)$ as $|q|\to\infty$ is
indeed not worse than powerlike (with the power depending on $M$). For a
Lorentz-covariant generalized function, the estimate $(u^j_M*g)(q)$ can be
given similarly using transformation rule~\thetag{29}, which leads to the
same conclusion, because matrix elements of the representation
$T^j_k(\Lambda)$ are rational functions of the boost parameter $q_+$.
Theorem~8 is proved.

\section{The decomposition into polynomial covariants}

We use the notation $(r,s)$, with nonnegative integer or half-integer $r$ and
$s$, for irreducible finite-dimensional representations of the $SL(2,\C)$
group and realize these representations in the ordinary way in the spaces of
complex homogeneous polynomials of the respective degrees $2r$ and $2s$ in
the spinor variables $\omega=(\omega_1, \omega_2)$ and $\ol\omega=
(\ol\omega_1,\ol\omega_2)$. We recall that the standard polynomial covariant
transforming according to the $(s,s)$ representation is given by $(\ol\omega\tl
x\omega)^{2s}$, where
\begin{equation}
\tl x= \begin{pmatrix} x_0-x_3&-x_1+ix_2 \\
x_1+ix_2&x_0+x_3\end{pmatrix}. \label{36}
\end{equation}
We now show that the representation~\cite{2},~\cite{23} of Lorentz-covariant
tempered distributions through polynomial covariants can be extended to
functionals of the class $S^{\pr\beta}$. A vector-valued generalized
function $w$ is now treated as a complex-valued generalized function in
the variable $x$, which in addition polynomially depends on the variables
$\omega$ and $\ol\omega$, and Lorentz-covariance condition~\thetag{28} becomes
\begin{equation}
w(x;\omega,\ol\omega)=
w\bigl(\Lambda(A)x;A\omega,\bar A\ol\omega\bigr),\quad A\in SL(2,\C),
\label{37}
\end{equation}
where $A\to\Lambda(A)$ is the canonical homomorphism of the $SL(2,\C)$ group
onto $L^{\uparrow}_+$.

{\bf Theorem 9.}
{\it The Lorentz-covariant generalized function $w$ that is defined on the
space $S^{\beta}$, $\beta\geq0$, and transforms according to the
representation $(r,s)$ is different from zero only if $r=s${\rm;} in this
case, it can be represented as
 \begin{equation}
w(x;\omega,\ol\omega)=(\ol\omega\tl x\omega)^{2s}v(x),
\label{38}
\end{equation}
where $v\in S^{\pr\beta}$ is a Lorentz-invariant functional determined by $w$
up to the term $\sum_{l=0}^{2s-1}c_l\square^l\delta(x)$ involving arbitrary
constants $c_l$.}

(We note that $(s,s)$ is a single-valued representation of the
$L^{\uparrow}_+$ group.)

Proving this theorem requires two lemmas.

{\bf Lemma 2.}
{\it Let $f\in S^{\beta}(\R^n)$, $\beta\geq0$. If $f|_{x_1=0}=0$, then $f(x)=
x_1f_1(x)$, where the function $f_1$ also belongs to $S^{\beta}(\R^n)$.}

{\bf Proof.}
We use the notation $x'=(x_2,\dots,x_n)$ and set
\begin{equation}
f_1(x_1,x')=\int_0^1(\ptl_1f)(tx_1,x')\,dt.
\label{39}
\end{equation}
For $|x_1|\leq1$, we have the estimate
\begin{equation}
\split \bigl|\ptl^{\kappa}f_1(x)\bigr|&\leq
\|f\|_{B,N}B^{|\kappa|+1}\frac{(\kappa_1+1)^{\beta(\kappa_1+1)}}
{\kappa_1^{\beta\kappa_1}}\kappa^{\beta\kappa}\bigl(1+|x'|\bigr)^{-N}\leq
\\
&\leq C_{\eps,N}\|f\|_{B,N}
(B+\eps)^{|\kappa|}\kappa^{\beta\kappa}\bigl(1+|x|\bigr)^{-N},
\endsplit
\label{40}
\end{equation}
where the norm is defined by Eq.~\thetag{2} and $\eps>0$ can be taken
arbitrarily small. For $|x_1|>1$, direct application of the Leibnitz formula
to $f_1=f/x_1$ gives
\begin{equation}
\bigl|\ptl^{\kappa}f_1(x)\bigr|\leq\|f\|_{B,N}\sum_{\ell\leq\kappa}
\binom{\kappa}{\ell}B^{|\kappa-\ell|}
(\kappa-\ell)^{\beta(\kappa-\ell)}\ell!\,\bigl(1+|x|\bigr)^{-N}.
\label{41}
\end{equation}
If $\beta>1$, then $\ell!\leq C_{\eps}\eps^{|\ell|}\ell^{\beta\ell}$, and
because of the inequality $(\kappa-\ell)^{\beta(\kappa-\ell)}\ell^{\beta\ell}
\leq\kappa^{\beta\kappa}$, we conclude that an estimate of type~\thetag{40}
is also valid for the function $f_1$ in this domain, i.e., $f_1\in S^{\beta}$.
We next let $\beta=1$. The space $S^1$ can be represented as a union over
$B$ of the spaces of functions that are analytic in the domains $T^B=
\bigl\{z=x+iy\in \C^n\colon |y_j|<1/B\ \forall j\bigr\}$ and have the finite
norms
\begin{equation}
\|f\|_{B,N}=\sup_{z\in T^B}\bigl|f(z)\bigr|\bigl(1+|x|\bigr)^N,\quad
N=1,2,\ldots.
\label{42}
\end{equation}
The condition $f|_{z_1=0}=0$ implies that $f(z)=z_1f_1(z)$, where $f_1$ is an
analytic function in the same tube as $f$ and is majorized by $f$ for
$|z_1|\geq\delta>0$. To estimate $f_1(z)$ for $|z_1|<\delta$, we use the
Cauchy formula with respect to $z_1$, setting $\delta=1/(3B)$ and taking a
circle of radius $2\delta$ as the integration contour. This contour lies in
$T^B$, and $|\zeta_1|>\delta$ for any point $\zeta$ belonging to it. As a
result, we obtain
\begin{equation}
\bigl|f_1(z)\bigr|\leq C_{\delta,N}\|f\|_{B,N}\bigl(1+|x|\bigr)^{-N}.
\label{43}
\end{equation}
For $\beta<1$, the estimate can be obtained similarly. In this case, $f_1(z)$
is an entire function, the norms are given by
\begin{equation}
\sup_z\bigl|f(z)\bigr|\bigl(1+|x|\bigr)^N\exp\bigl(-|By|^{1/(1-\beta)}\bigr),
\label{44}
\end{equation}
and $\delta$ can be set equal to $1$. Lemma~2 is thus proved.

We note that the occurrence of the norm of $f$ in right-hand sides of the
estimates demonstrates the sequential continuity of the mapping that is
inverse to the injective mapping $f\to x_1f$ of $S^{\beta}$ into itself.

{\bf Lemma 3.}
{\it Any function $f\in S^{\beta}(\R^n)$, $\beta\geq0$, satisfying the
condition $\ptl^{\kappa}f(0)=0$ for all $|\kappa|\leq mn$ admits a
decomposition of the form}
 \begin{equation}
f(x)=\sum_{i=1}^nx_i^{m+1}f_i(x),\quad\text{where }f_i\in S^{\beta}(\R^n).
\label{45}
\end{equation}

{\bf Proof.}
For $n=1$, this representation directly follows from Lemma~2. We next use
induction on $n$ with the notation
\begin{equation}
g_j(x')=\frac{1}{j!}\ptl^j_1f(x)\bigr|_{x_1=0},\qquad
F(x)=f(x)-f_0(x_1)\sum_{j=0}^mx^j_1g_j(x'),
\label{46}
\end{equation}
where $x'=(x_2,\dots,x_n)$ as before and the function $f_0\in S^{\beta}(\R)$
is subjected to the conditions $f_0^{(j)}(0)=0$, $0\leq j\leq m$. For all
$|\kappa|\leq m(n-1)$, we have $\ptl^{\kappa}g_j(0)=0$, and by the induction
hypothesis,
\begin{equation}
g_j(x')=\sum_{i=2}^nx_i^{m+1}f_{ij}(x'),\quad\text{where }
f_{ij}\in S^{\beta}(\R^{n-1}).
\label{47}
\end{equation}
Next, $\ptl^j_1F|_{x_1=0}=0$ for all $j\leq m$. Therefore, $F(x)=x_1^{m+1}
f_1(x')$ by Lemma~2, and Eq.~\thetag{45} is satisfied with $f_i(x)=f_0(x_1)
\sum_{j=0}^mx_1^jf_{ij}(x')$, $2\leq i\leq n$.

{\bf Proof of Theorem 9.}
The identical vanishing of $w$ for $r\neq s$ follows from Theorem~8 and from
Proposition~3.6 in~\cite{2} describing the structure of the Lorentz-covariant
Schwartz distributions. The dimension of the representation $(s,s)$, i.e.,
the number of different monomials in $\omega$ and $\ol\omega$ of the degree
$2s$ in each of these variables, is given by $(2s+1)^2$. We enumerate these
monomials and consider the mapping
\begin{equation}
h\:\bigoplus^{(2s+1)^2}S^{\beta}\to S^{\beta}
\nonumber
\end{equation}
that takes each set of test functions $\{f_i\}$, $1\leq i\leq(2s+1)^2$ to
their linear combination obtained by replacing the monomials in the polynomial
$(\ol\omega\tl x\omega)^{2s}$ with the test functions with the corresponding
indices. Let $h'$ denote the dual mapping of $h$. To each functional $v\in
S^{\pr\beta}$, it assigns the set of functionals obtained by multiplying $v$
by the coefficients of the polynomial, and the restriction of $h'$ to
the subspace of invariant functionals takes them into covariant functionals
of form~\thetag{38}. Because any covariant Schwartz distribution has this
form and these distributions are dense in covariant generalized functions of
the class under consideration, it suffices to show that the image of any
closed subspace under $h'$ is closed in $\bigoplus^{(2s+1)^2}S^{\pr\beta}$.

We note that $\Ima h$ contains a closed subspace of finite codimension in
$S^{\beta}$, specified by the conditions $\ptl^{\kappa}f(0)=0$,
$|\kappa|\leq4s(4s-1)$. Indeed, by Lemma~3, we have the decomposition
$f(x)=\sum_{i=0}^3x_i^{4s}f_i(x)$; with $x_i^{4s}$ expressed through the
elements $x_{\rho\sigma}$ of matrix~\thetag{36}, each term of the resulting
expression contains at least one of these elements raised to a power not less
than $2s$. Therefore, the function $f$ can be written as the sum
\begin{equation}
\sum_{\rho,\sigma=1,2}x_{\rho\sigma}^{2s}f_{\rho\sigma},\quad
f_{\rho\sigma}\in S^{\beta},
\nonumber
\end{equation}
which is obviously in $\Ima h$. The subspace $\Ima h$ can therefore be
represented as a sum of a finite-dimensional and a closed subspace and is
therefore closed. Next, because $S^{\beta}$ is reflexive, the closedness of
$\Ima h$ implies that $h'$ is a homomorphism (an open mapping onto its image)
with respect to the strong topology of the dual spaces~\cite{43} (Sec.~4.4.1).
Let $L$ be a closed subspace of $S^{\pr\beta}$. Its sum with the
finite-dimensional subspace $\Ker h'$ (which is contained in the linear span
of the functionals $\ptl^{\kappa}\delta(x)$, $|\kappa|\leq4s(4s-1)$, is also
closed. Therefore, for a point $v\notin L+\Ker h'$, there exists a
neighborhood $\U$ that does not intersect $L+\Ker h'$. The set $h'(\U)$ is a
neighborhood in $\Ima h'$ and does not intersect the subspace $h'(L)$.
Therefore, $h'(L)$ is closed, as was to be proved.

The last assertion in the theorem defines the kernels of the restriction of
$h'$ to the subspace of Lorentz-invariant functionals more exactly. Because
the invariant combinations of the distributions $\ptl^{\kappa}\delta(x)$ are
of the form $\sum_lc_l\square^l\delta(x)$ and are converted into polynomials
in $p^2$ by the Fourier transformation and because the Fourier transform of
$(\ol\omega\tl x\omega)$ is $-i(\ol\omega\tl\ptl\omega)$, the proof is
completed by applying the identities $(\ol\omega\tl\ptl\omega)(p^2)=
2(\ol\omega\tl p\omega)$ and $(\ol\omega\tl\ptl\omega)(\ol\omega\tl p\omega)=
0$. This implies that $(\ol\omega\tl\ptl\omega)^{2s}(p^2)^l=0$ only for
$l\leq2s-1$. Theorem~9 is proved.

We can now extend Theorem~6 to covariant generalized functions, but this
requires one more auxiliary statement, which is closely related to Lemma~2.

{\bf Lemma 4.}
{\it Let $v\in S^{\pr\beta}$, $\beta\leq1$, and let $K$ be a closed cone in
$\R^n$ that contains the plane $x_1=0$. If $K$ is a carrier cone of $x_1v$,
then it is also a carrier of $v$.}

{\bf Proof.}
Let $O$ be the union of an open cone $U\supset K\setminus\{0\}$ and an
$\eps$-neighborhood of the origin, and let $f\in S^{\beta}(O)$. We set
$g(z')=f(0,z')$. Because the points $(0,z')$ lie in $O$, the function $g$
belongs to $S^{\beta}(\R^{n-1})$, and
\begin{equation}
\|g\|_{B,N}\leq\|f\|_{O,B,N}.
\label{48}
\end{equation}
For $\beta=1$, the norm $\|f\|_{O,B,N}$ is defined differently than
in~\thetag{42}; namely, $\sup$ is now taken over the complex $(1/B)$-neighborhood
of $O$, while for $\beta<1$, the norm involves the factor
$\exp\bigl\{-d(Bx,U)^{1/(1-\beta)}\bigr\}$ in addition to~\thetag{44} in
accordance with Eq.~\thetag{13}. Let $f_0(z_1)$ be any function belonging to
$S^{\beta}(\R)$ that is equal to~1 at the origin. Then
$(f-f_0g)\in S^{\beta}(O)$ and $(f-f_0g)\bigr|_{z_1=0}=0$. The same
elementary argument as in the proof of Lemma~2 shows that $(f-f_0g)(z)=
z_1f_1(z)$, where $f_1$ belongs to $S^{\beta}(O)$ and tends to zero in this
space as $f$ tends to zero in its topology. Therefore, the
formula
\begin{equation}
(\hat v,f)=(v,f_0g)+(x_1v,f_1)
\label{49}
\end{equation}
defines a continuous extension of the functional $v$ to $S^{\beta}(O)$, which
proves the statement of the lemma. A similar statement is also true for
$\beta>1$, but it is trivial in that case. Lemma 4 is proved.

{\bf Theorem 10.}
{\it Any Lorentz-covariant generalized function $w$ over $S^{\beta}$,
${\beta\geq0}$, with the carrier cone $\ol\V$ admits a decomposition into
Lorentz-covariant generalized functions of the same class with the carrier
cones $\ol\V_+$ and $\ol\V_-$.}

{\bf Proof.}
We assume that $w$ transforms according to the $(s,s)$ representation. The
cone $\ol\V$ is a carrier of the invariant functional $v$ through which $w$
is expressed by Eq.~\thetag{38}. Indeed, by the condition of the theorem, it
is a carrier of $(x_0-x_3)^{2s}v$. Viewing the difference $x_0-x_3$ as the
first coordinate and applying Lemma~4, we conclude that the complement of a
conical neighborhood of the positive $x_3$ semiaxis is certainly a carrier
cone of $v$, and in view of the Lorentz invariance, the same is true for any
other spacelike direction; the intersection of these complementary cones is
exactly $\ol\V$. It remains to take the existence of the smallest carrier
cone into account and apply Theorem~6. Theorem~10 is proved.

\section{Application to the spin--statistics theorem}

We consider a finite set of fields $\{\phi_i\}$ that are operator-valued
generalized functions over the space $S^{\beta}(\R^4)$, $\beta<1$, and satisfy the
standard assumptions of the Wightman axioms~\cite{1},~\cite{2} except for the
local commutativity, which is impossible to formulate using analytic test
functions. A natural replacement for this axiom, with its meaning being
closer to the physical requirement of macrocausality, is the condition that
the closed light cone $\ol\V$ be a carrier cone of the matrix elements of the
commutators of observable fields. (If the theory also involves nonobservable
fields, then $\ol\V$ is a carrier cone of either commutators or
anticommutators. For more details on the motivation and the exact
formulation of this condition, which we call asymptotic commutativity,
see~\cite{24}.) We follow the standard assumption that the commutation
relation type depends on only the type of the field and is the same for all
of its Lorentz components; therefore, Lorentzian indices can be omitted in
what follows. From the transformation properties of the fields under the
Poincar\'e group and the invariance of the vacuum, it follows that the vacuum
expectation values $\bigl\langle\Psi_0,\phi(x_1)\psi(x_2)\Psi_0\bigr\rangle$
are Lorentz-covariant generalized functions over $S^{\beta}(\R^4)$ with respect to
the difference variable $\xi=x_1-x_2$. This allows applying Theorem~9 to
generalize the derivation of the spin--statistics relation to nonlocal fields. In
complete analogy with the standard theory of tempered quantized
fields~\cite{1},~\cite{2}, the weak cluster decomposition property of vacuum expectation
values (which follows from the existence and uniqueness of the vacuum without
using locality) implies that any pair of nonzero fields $\phi$, $\psi$
defined on $S^{\beta}$ has commutation relations of the same type as the pair
$\phi$, $\psi^*$ (see Theorem~11 in~\cite{24}). Therefore, the problem
reduces to the analysis of asymptotic commutation relations between the field
$\phi$ and its Hermitian conjugate field $\phi^*$.

{\bf Theorem 11.}
{\it Let $\phi$ be a field defined on the space $S^{\beta}(\R^4)$,
$0\leq\beta<1$, transforming according to the irreducible representation
$(r,s)$ of the $SL(2,\C)$ group. The anomalous asymptotic commutation
relation between $\phi$ and $\phi^*$ {\rm(}anticommutativity for an integer
spin and commutativity for a half-integer spin{\rm)} implies the equality
$\phi(f)\Psi_0=\phi^*(f)\Psi_0=0$ for all $f\in S^{\beta}(\R^4)$.}

{\bf Proof.}
We use the notation
\begin{equation}
W(x_1-x_2)=\bigl\langle\Psi_0,\phi(x_1)\phi^*(x_2)\Psi_0\bigr\rangle,\qquad
W'(x_1-x_2)=\bigl\langle\Psi_0,\phi^*(x_1)\phi(x_2)\Psi_0\bigr\rangle
\label{50}
\end{equation}
and first consider the scalar field case. The anomalous asymptotic
commutation relation implies that the cone $\ol\V$ is a carrier of the
functional $W(\xi)+W'(-\xi)$. In accordance with Theorem~7, this functional
is then odd because by the spectral condition, its Fourier transform vanishes for
$p^2<0$. Therefore, the functional $W(\xi)+W'(\xi)$ is also odd; in
momentum space, this functional is supported by the cone $\ol\V_+$
and must then be identically equal to zero. Taking its value on a test
function of the form $\bar f(x_1)f(x_2)$, we obtain
\begin{equation}
\bigl\|\phi(f)\Psi_0\bigr\|^2+\bigl\|\phi^*(f)\Psi_0\bigr\|^2=0,\quad
f\in S^{\beta}.
\label{51}
\end{equation}

We now let the field $\phi$ transform according to the irreducible representation
$(r,s)$. The anomalous commutation relation then implies that $\ol\V$ is a
carrier cone of $W(\xi)\pm W'(-\xi)$, where the plus sign corresponds to an
integer spin case and the minus sign to a half-integer one. Lorentz-covariant
generalized functions~\thetag{50} transform according to the representation
$(r,s)\otimes(s,r)$, whose decomposition into irreducible representations is
given by
\begin{equation}
(r,s)\otimes(s,r)=\bigoplus_{\stackrel
{|r-s|\leq r',s'\leq r+s}{r',s'\in|r-s|+\Bbb N}}(r',s').
\label{52}
\end{equation}
Accordingly, the decomposition of $W(\xi)\pm W'(-\xi)$ into polynomial
covariants then involves $2\min(r,s){+}1$ terms. We note that in the integer
spin case, where $r+s$ is an integer, the decomposition involves covariants
of only even degrees, while for the half-integer spin, only odd. We apply
Theorem~10 and perform the Laplace transformation. In momentum space,
the distribution $\wtl W(p)\pm \wtl W^{\,\pr}(-p)$ is then represented as the
difference of boundary values of Lorentz-covariant analytic functions that
are holomorphic in the tubes $\Bbb T_{\pm}$ and can be analytically continued
to the extended domain $\Bbb T^{\ext }$ by the Bargmann--Hall--Wightman
theorem~\cite{1},~\cite{2}. These analytic functions are symmetric with
respect to the full reflection $p+i\eta\to-p-i\eta$ for an integer spin and
are antisymmetric for a half-integer spin, and in view of the spectral condition and
the uniqueness theorem, they coincide with each other in $\Bbb T^{\ext }$.
For the boundary values, this then implies the identity
\begin{equation}
\wtl W(p)\pm\wtl W'(-p)=\mp\wtl W(-p)-\wtl W'(p).
\label{53}
\end{equation}
Again taking the spectral condition into account, we see that only the point
$p=0$ can be the support of $\wtl W(p)+\wtl W^{\,\pr}(p)$. The singularity
order of this distribution must be equal to zero by the positivity condition.
In the case with a half-integer spin, there is no such term in its decomposition into
covariants, while for an integer spin, it cannot transform in accordance with
Eq.~\thetag{53} under the reflection. Therefore, $\wtl W(p)+\wtl W^{\,\pr}(p)
\equiv0$, which completes the proof.

\section{Concluding remarks}

The results obtained here allow treating highly singular Lorentz-covariant
generalized functions as easily as the standard tempered distributions. An essential
addition to Theorem~7 is given by Theorem~2.14 in~\cite{22} and Theorem~9
in~\cite{24}, which show that the cone $\ol\V$ is a carrier cone of odd
Lorentz-invariant functionals with arbitrarily singular behavior. In Sec.~9, we
considered two-point Wightman functions of a special form; however, there is
a natural analogue of the covariant decomposition~\cite{2},~\cite{23} for the
vacuum expectation values of any pair of fields over $S^{\beta}$ that
transform according to finite-dimensional irreducible or simply reducible $SL(2,\C)$
representations. Among open problems, we mention the proof of an analogue of
the Meth\'ee representation for the even and odd invariant functionals of
class $S^{\pr\beta}$. We also note that the theorems proved above can be extended to
functionals defined on the generalized Gelfand--Shilov spaces $S^b$ defined
by an indicator function $b$ characterizing the growth of their Fourier
transforms. The corresponding restrictions on the indicator function are
established in~\cite{21}.

{\bf  Acknowledgments}
This work is supported by the Russian Foundation for Basic Research (Grant
Nos.~99-01-00376 and~00-15-96566).

\end{document}